\documentclass[12pt]{iopart}
\usepackage{iopams}  
\usepackage{graphicx}
\usepackage{color}
\usepackage{hyperref}
\usepackage{etoolbox}

\begin{document}

\title[]{Valley filters using graphene blister defects from first principles}

\author{Mitsuharu Uemoto, Masaki Nishiura, Tomoya Ono}

\address{Department of Electrical and Electronic Engineering, Graduate School of Engineering, Kobe University, Nada, Kobe, 657-8501 Japan}
\ead{uemoto@eedept.kobe-u.ac.jp}
\vspace{10pt}
\begin{indented}
\item[] \today
\end{indented}

\begin{abstract}
Valleytronics, which makes use of the two valleys in graphenes, attracts considerable attention and a valley filter is expected to be the central component in valleytronics. We propose the application of the graphene valley filter using blister defects to the investigation of the valley-dependent transport properties of the Stone--Wales and blister defects of graphenes by density functional theory calculations. It is found that the intervalley transition from the $\mathbf{K}$ valley to the $\mathbf{K}^\prime$ valleys is completely suppressed in some defects. Using a large bipartite honeycomb cell including several carbon atoms in a cell and replacing atomic orbitals with molecular orbitals in the tight-binding model, we demonstrate analytically and numerically that the symmetry between the A and B sites of the bipartite honeycomb cell contributes to the suppression of the intervalley transition. In addition, the universal rule for the atomic structures of the blisters suppressing the intervalley transition is derived. Furthermore, by introducing  additional carbon atoms to graphenes to form blister defects, we can split the energies of the states at which resonant scattering occurs on the $\mathrm{K}$ and $\mathrm{K}^\prime$ channel electrons. Because of this split, the fully valley-polarized current will be achieved by the local application of a gate voltage.
\end{abstract}

\section{Introduction}
Graphene, a two-dimensional material with a bipartite honeycomb cell (BHC), has been the subject of extensive research as a material of next-generation electronics \cite{ Science.32.5934, AdvMat.24.43}. The valence and conduction bands of the graphene exhibit a characteristic band structure called the Dirac cone. In the Brillouin zone, there are two inequivalent Dirac cones represented as $\mathbf{K}$ and $\mathbf{K}^\prime$, which provide additional binary degrees of freedom for the transport carriers known as the valley index. The valleys are separated by a large wave vector and the valley information is robust against intervalley transition from slowly varying potentials. This characteristic feature of the valley degrees of freedom opens the possibility that the valley degrees of freedom might be used to control an electronic device in the same way with the electron spin, which is used in spintronics or quantum computing. The extension of electronics using valley degrees of freedom is called ``valleytronics'' \cite{Small.14.1801483}.

Generation, detection, and control of valley polarization are one of the most important techniques for realization of valleytronics devices. For example, valley-dependent optical selection rules enable conversions between light field and valley polarization \cite{NatPhoton.12.451, PhysRevB.105.115403}. In addition, the valley filter is central component in valleytronics devices, just as the spin filter is central in spintronics devices \cite{Small.14.1801483,NatPhys.3.172,PhysRevLett.106.136806,PhysRevLett.117.276801,Phys.Rev.Appl.11.044033,Phys.Rev.B.99.064302, Carbon.160.353, Phys.Rev.B.107.035411,JPhysCondMatt.32.365302,tian2022valleytronics,ingaramo2016valley}. In graphene, the BHC with A and B sites enforces the Berry curvature of $\mathbf{K}$ into $\mathbf{K}^\prime$ and renders them indistinguishable. Several methods introducing atom-scale structure deformations and local strains that break the symmetry of BHCs \cite{NanoLett.17.5304} have been proposed so far for the graphene-based valley filters. Examples include artificial nanostructures such as quantum point contact devices \cite{NatPhys.3.172}, lattice defects like grain boundaries and line defects  \cite{PhysRevLett.106.136806, Phys.Scr.97.125825, JPhysCondMatt.32.365302,tian2022valleytronics,ingaramo2016valley}, and graphene nano blisters \cite{PhysRevLett.117.276801}. However, practical and effective methods to manipulate the valleys in realistic setups still need to be established. Graphene blisters are created via electron beam irradiation \cite{NanoLett_14_908} and exhibit out-of-plane deformation different sizes and shapes by adding a couple of carbon atoms on graphene. Since their structures are simple as well as Stone--Wales (SW) defects, it will be instructive to make some general symmetry analysis for the valley filter effect. Atomistic tight-binding models have predicted the valley-dependent transport property; this suggests that the breaking of atomic-scale inversion symmetry results in an intervalley transition, which is essential for inducing valley polarization \cite{NJPhys.18.113011, Phys.Scr.97.125825}. 

In this study, we propose the use of blister defects suitable for valley filters by investigating valley-dependent transport properties in the graphene with SW and blister defects using density functional theory (DFT) calculations \cite{PhysRev_136_B864}. It is found that the intervalley transition is completely suppressed, although BHCs are distorted in some defects. We prove analytically and numerically the relationship between the local structure of the defects and the suppression of the intervalley transition by introducing a large BHC (LBHC) containing several atoms in one cell and replacing atomic orbitals with molecular orbitals in the tight-binding model. 
To the best of our knowledge, there have been no reports thus far on the universal rules to control the intervalley transition in the graphene blister defects from continuum descriptions like DFT calculations, and we provide the rule for the atomic structure of the blister defects for the valley filter. 
Moreover, by adding the extra carbon atoms to form the blisters, one can split the energies of the states where the resonant scattering occurs on the $\mathrm{K}$ and $\mathrm{K}^\prime$ channel electrons. This split will enable us to realize fully valley-polarized currents by local application of a gate voltage to the blister region. 

This paper is organized as follows: Section~\ref{sec:Method} presents the details of our computational conditions and defect structures by the DFT calculation. In Sec.~\ref{sec:Results and discussion}, we report the valley-dependent transport properties of the proposed structures and provide an analytical and numerical demonstrations for the suppression of the intervalley transition in the blisters. In addition, we propose the universal rule to control the intervalley transition. Finally, in Sec.~\ref{sec:conclusion}, we conclude our study. In \ref{sec:tight_binding}, the numerical formalism of the tight-binding model is introduced.

\section{Method}
\label{sec:Method}
\begin{figure}[htb]
\begin{center}
\includegraphics{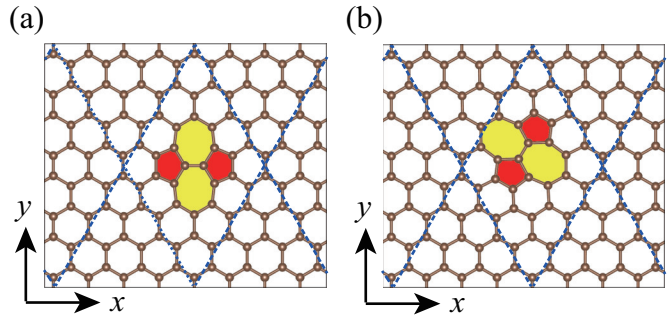}
\caption{Atomic structures of scattering regions of (a) S$_1$ and (b) S$_2$ structures. Blue dotted lines indicate the LBHC containing several carbon atoms.}
\label{fig:1}
\end{center}
\end{figure}

For the DFT calculation, we employ the RSPACE code,\cite{PhysRevLett.82.5016, PhysRevB.72.085115, KikujiHirose2005, PhysRevB.82.205115} which uses the real-space finite-difference approach \cite{PhysRevLett_72_001240, PhysRevB_50_011355} within the frameworks of DFT and enables us to investigate transport properties owing to the high degree of freedom for boundary conditions. As an example, we show in Fig.~\ref{fig:1}(a) the computational models of the scattering region, in which the SW defect is embedded and electrons are injected along the positive direction of the $x$ axis. First, we examine the optimized atomic structures of graphene without any defects as well as that with a defect using supercells under the conventional periodic boundary condition. The computational region is set to be 19.68 $\times$ 17.04 $\times$ 10.58 \AA$^3$ with 128 $\times$ 80 $\times$ 50 grid points, and a 6 $\times$ 6 Monkhorst--Pack $k$-point mesh including a $\Gamma$-point is taken in the Brillouin zone. The exchange-correlation effect is treated by the local density approximation \cite{PhysRevB_23_5048} of the DFT, and the projector augmented wave method \cite{PhysRevB_50_17953} is used to describe the electron-ion interaction. We then examine the valley-dependent transport properties of defect structures by attaching graphene electrodes to both the left and right sides of the scattering region. In the transport-property calculation performed to obtain the scattering wave functions, the norm-conserving pseudopotentials \cite{ComputMaterSci_14_72} of Troullier and Martins \cite{PhysRevB_43_1993} are employed. To determine the Kohn--Sham effective potential, a supercell is used under a periodic boundary condition, and then the scattering wave functions are computed under the semi-infinite boundary condition non-self-consistently. It has been reported that this procedure is just as accurate in the linear response regime but significantly more efficient than performing computations self-consistently on a scattering-wave basis \cite{PhysRevB_76_235422}. Integration over the Brillouin zone is carried out using the $\Gamma$ point for the scattering region, and the $k$-point mesh in the electrode regions is chosen to correspond to that in the scattering region. The relationship between the valley-dependent transport property and the local atomic structures of the defects is investigated.

\section{Results and discussion}
\label{sec:Results and discussion}
\subsection{SW defects}
\label{subsec:SW defects}
\begin{figure}[htb]
\begin{center}
\includegraphics{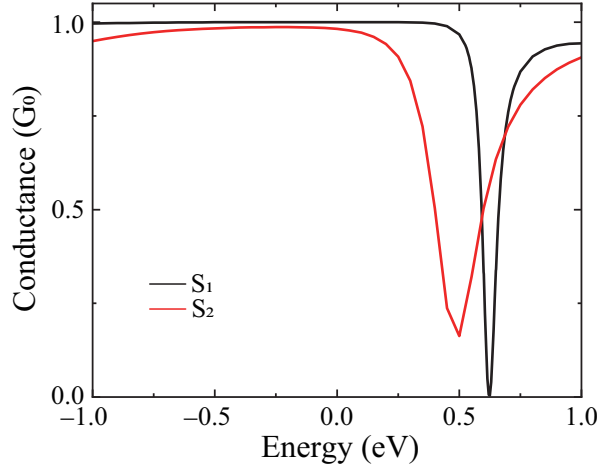}
\caption{Conductance spectra of SW defects.}
\label{fig:2}
\end{center}
\end{figure}

\begin{figure}[htb]
\begin{center}
\includegraphics{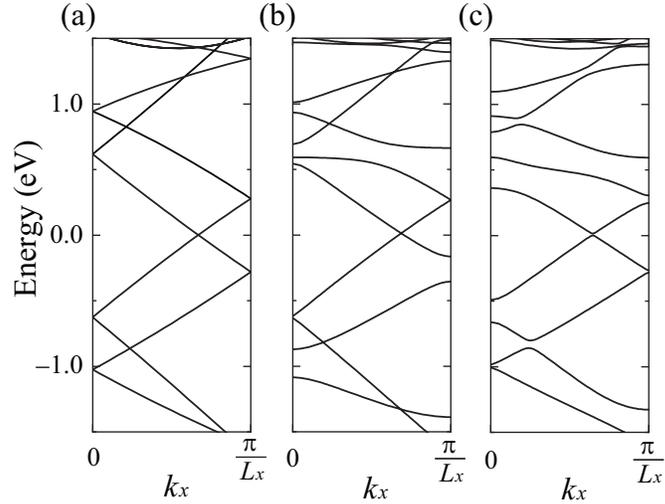}
\caption{Electronic band structures of (a) pure graphene, (b) S$_1$, and (c) S$_2$ structures.}
\label{fig:3}
\end{center}
\end{figure}

\begin{figure*}[htb]
\begin{center}
\includegraphics[width=0.75\textwidth]{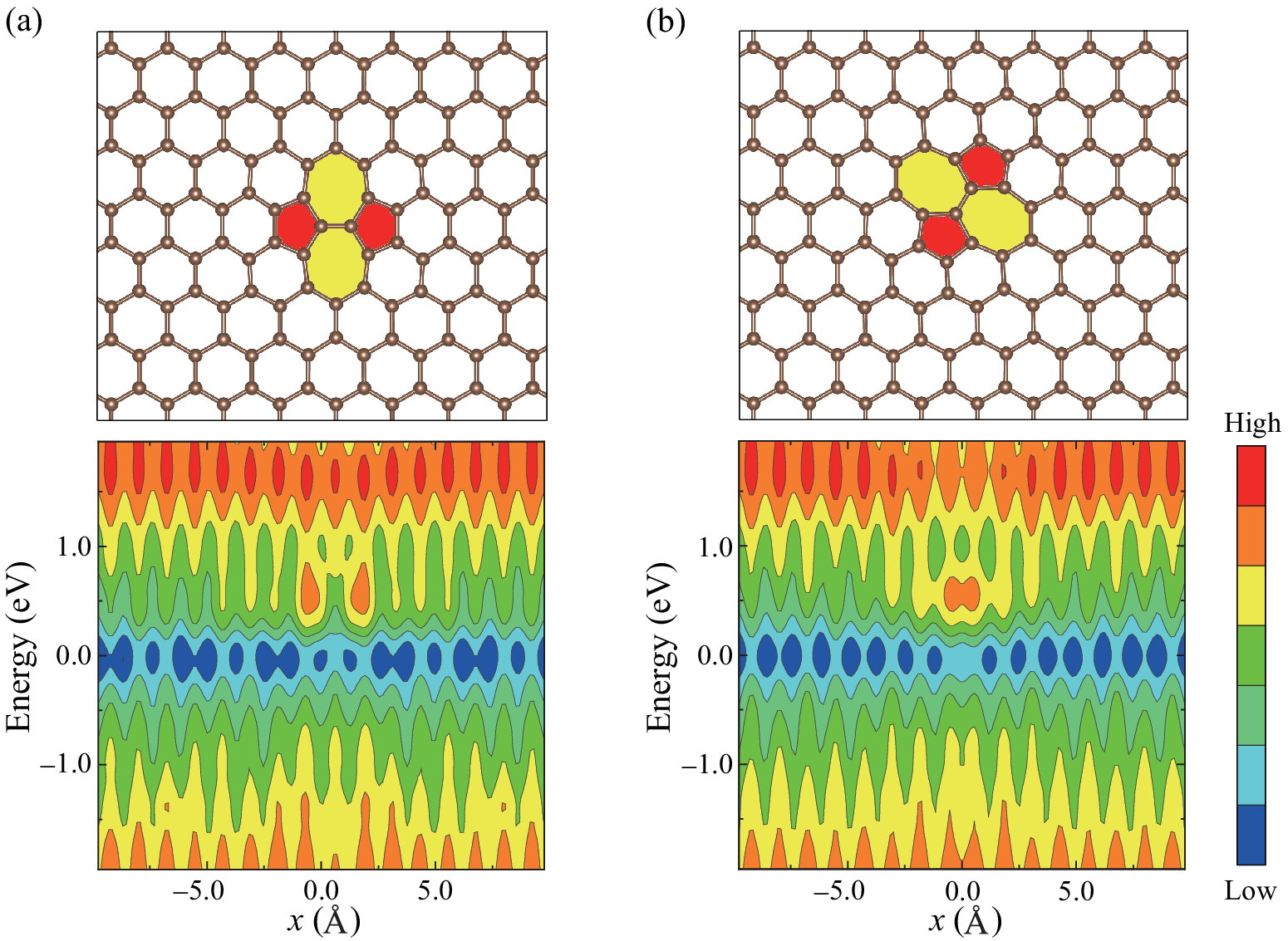}
\caption{LDOSs of (a) S$_1$ and (b) S$_2$ structures. For clarity, structural models are provided on the top of each distribution. Zero energy is chosen as the Fermi level. Each contour represents twice or half the density of the adjacent contours and the lowest contour is 3.78 $\times$ 10$^{-3}$ electron/eV/\AA.}
\label{fig:4}
\end{center}
\end{figure*}

\begin{figure}[htb]
\begin{center}
\includegraphics{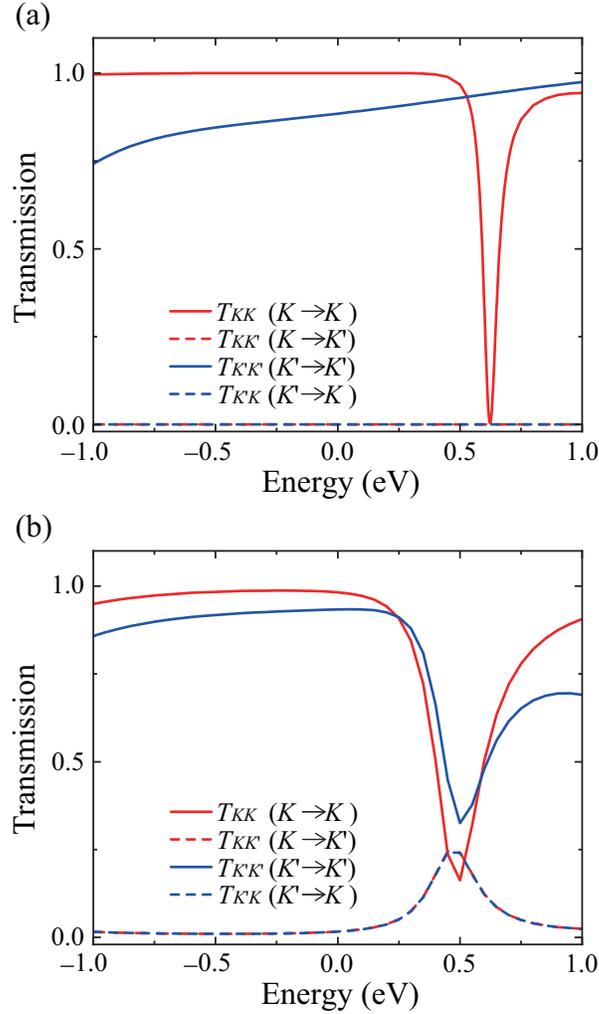}
\caption{Channel-resolved transmission spectra of (a) S$_1$ and (b) S$_2$ structures.}
\label{fig:5}
\end{center}
\end{figure}

SW defects involve an in-plane $\pi/2$ rotation of the C--C bond with respect to the midpoint of the bond. In this study, we prepare two computational structures of SW defects, where the direction and location of the rotated C--C bond are changed. In Fig.~\ref{fig:1}(a), the rotated C--C bond is parallel to the $x$ axis, while in Fig.~\ref{fig:1}(b), the rotated C--C bond forms an angle of $-\pi$/6 to the $x$ axis. The structures shown in Figs.~\ref{fig:1}(a) and \ref{fig:1}(b) are named S$_1$ and S$_2$, respectively. The total conductance is shown in Fig.~\ref{fig:2}, in which the conductance shows a dip at $E_F + 0.5$~eV in both the S$_1$ and S$_2$ structures. Figure~\ref{fig:3} shows the electronic band structures of graphene and SW defects. Due to the insertion of the SW defects, the electronic band structures at around $E_F + 0.5$~eV  are modified. We plot the local density of states (LDOS), which is calculated as
\begin{equation}
\rho(x,E)=\sum_{l,k} \int |\Psi_{l,k}(x,y,z)|^2 \mathrm{d}y \mathrm{d}z \times N \mathrm{e}^{-\alpha(E-\varepsilon_{l,k})^2},
\end{equation}
where $\varepsilon_{l,k}$ are the eigenvalues of the wave function $\Psi_{l,k}$ with indexes $l$ and $k$ denoting the eigenstate and the $k$-point, respectively, $x$ is the coordinate of the plane where the LDOSs are plotted, and $N(=2\sqrt{\frac{\alpha}{\pi}})$ is the normalization factor with $\alpha$ as the smearing factor. Here, $\alpha$ is set to 13.5~eV$^{-2}$. According to the LDOSs plotted in Fig.~\ref{fig:4}, it is found that the dip can be ascribed to the resonant scattering by the defect states at the SW defects. The channel-resolved transmission spectra are plotted in Fig.~\ref{fig:5}. In general, the wave functions of a transmitted wave through the scattering region $\psi$ are written as a linear combination of propagating waves in the right electrode:
\begin{equation}
\psi=\sum_{l=1}^n t_l\phi_l^\mathrm{tra},
\end{equation}
where $\phi_l^\mathrm{tra}$s represent the propagating waves in the electrode regions, $t_l$s are the transmission coefficients, and $n$ is the number of propagating waves in the right electrodes. In the case of graphene, there are two propagating waves around the Fermi level, which belong to the $\mathbf{K}$ and $\mathbf{K}^\prime$ valleys. The channels of electrons in the $\mathbf{K}$ and $\mathbf{K}^\prime$ valleys are distinguished by their group velocities and wave vectors. Thus, the transmitted wave functions are described as
\begin{equation}
\psi=t_K\phi_K^\mathrm{tra}+t_{K^\prime}\phi_{K^\prime}^\mathrm{tra}.
\label{eq:separate}
\end{equation}
Even if electrons in the $\mathbf{K}$ valley are injected, the intervalley transition from the $\mathbf{K}$ valley to the $\mathbf{K}^\prime$ valley occurs at the scattering region in some cases. The transmission probabilities for incident electrons in the $\mathbf{K}$ and $\mathbf{K}^\prime$ valleys are plotted in Fig.~\ref{fig:5}. In the case of the S$_1$ structure, the transmission probability of the $\mathbf{K}$ channel is almost unity near the Fermi level, while that of the $\mathbf{K}^\prime$ channel is considerably reduced by the scattering at the SW defects. On the other hand, electrons in both the $\mathbf{K}$ and $\mathbf{K}^\prime$ channels are scattered at $E_F + 0.5$~eV. The intervalley transition from the $\mathbf{K}$ valley to the $\mathbf{K}^\prime$ valley does not occur in the S$_1$ structure, while it occurs in the S$_2$ structure at $\sim E_F+ 0.5$~eV. The tight-binding model \cite{JPhysSocJpn_67_2421} is employed to interpret the intervalley transition for the BHC of graphene in the next subsection.

\subsection{Analytical investigation for scattering at defects}
\label{sec:A1}
\begin{figure}[htb]
\begin{center}
\includegraphics{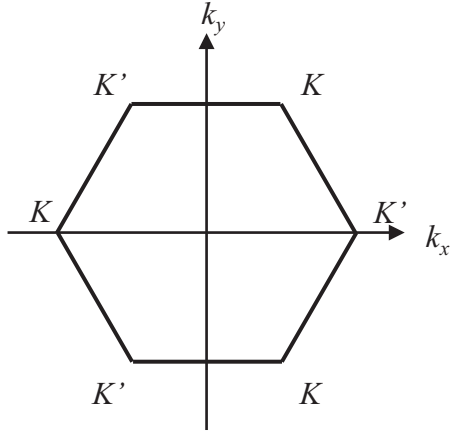}
\caption{Brillouin zone of pure graphene}
\label{fig:A1}
\end{center}
\end{figure}

Let us consider the $\pi$ band tight-binding Hamiltonian for graphene near the $\mathbf{K}[=(K_x, K_y)]$ point in the Brillouin zone shown in Fig.~\ref{fig:A1}

\begin{eqnarray}
\hat{\mathcal{H}}(\mathbf{K}+\mathbf{k}) \approx
\left(
\begin{array}{cc}
u_A & \gamma k \mathrm{e}^{ i \varphi(\mathbf{k})} \\
\gamma k \mathrm{e}^{- i \varphi(\mathbf{k})} & u_B
\end{array}
\right), \nonumber
\end{eqnarray}
where $\mathbf{k}=(k_x, k_y)$, $k=|\mathbf{k}|$, $\varphi(\mathbf{k})$ is an angle of the wave vector $\mathbf{k}$, and $\gamma$ is a band parameter. It is obvious that $u_A=u_B=u$ in the case of pristine graphene because the atoms at A and B sites are carbon. The eigenvalue and eigenvector of the Hamiltonian are, respectively, $\varepsilon_s^K = u \pm s \gamma k$ and
\begin{eqnarray}
\phi_s^K = \frac{1}{\sqrt{2}} \left(
\begin{array}{c}
s \\
\mathrm{e}^{ i \varphi(\mathbf{k})}
\end{array}
\right). \nonumber
\end{eqnarray}
Here, $s$ denotes the bands and the state with $s=+1$ ($s=-1$) corresponds to the conduction (valence) band. The group velocity is expressed as $v_s^K=s \gamma$.

In a similar manner, the Hamiltonian, eigenvalue, eigenvector, and group velocity near the $\mathbf{K}^\prime$ point are
\begin{eqnarray}
\hat{\mathcal{H}}(\mathbf{K}^\prime+\mathbf{k}) \approx
\left(
\begin{array}{cc}
u_A & \gamma k \mathrm{e}^{- i \varphi(\mathbf{k})} \\
\gamma k \mathrm{e}^{i \varphi(\mathbf{k})} & u_B
\end{array}
\right), \nonumber
\end{eqnarray}
$\varepsilon_s^{K^\prime} = u \pm s \gamma k$,
\begin{eqnarray}
\phi_s^{K^\prime} = \frac{1}{\sqrt{2}} \left(
\begin{array}{c}
\mathrm{e}^{ i \varphi(\mathbf{k})} \\
s
\end{array}
\right), \nonumber
\end{eqnarray}
and $v_s^{K^\prime}=s \gamma$, respectively. In addition, the envelop functions \cite{JPhysSocJpn_67_2421} are expressed as
\begin{eqnarray}
\mathbf{F}^K_{s\mathbf{k}}(\mathbf{r})=\frac{1}{N}\mathrm{e}^{i \mathbf{k} \cdot \mathbf{r}} \phi_s^K, \nonumber \\
\mathbf{F}^{K^\prime}_{s\mathbf{k}}(\mathbf{r})=\frac{1}{N}\mathrm{e}^{i \mathbf{k} \cdot \mathbf{r}} \phi_s^{K^\prime}, \nonumber
\end{eqnarray}
with $N$ being a normalization constant.

\begin{figure}[htb]
\begin{center}
\includegraphics{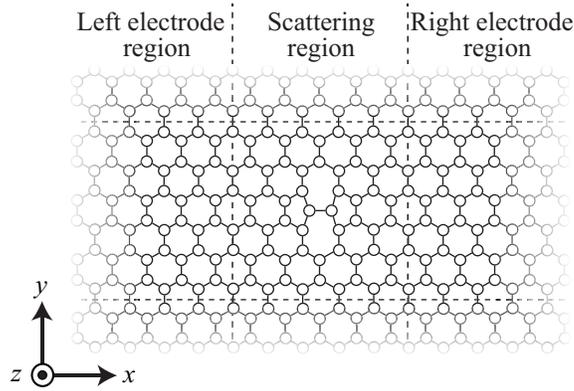}
\caption{Sketch of electrode and scattering regions for transport calculation. White circles indicate carbon atoms.}
\label{fig:A2}
\end{center}
\end{figure}

\begin{figure}[htb]
\begin{center}
\includegraphics{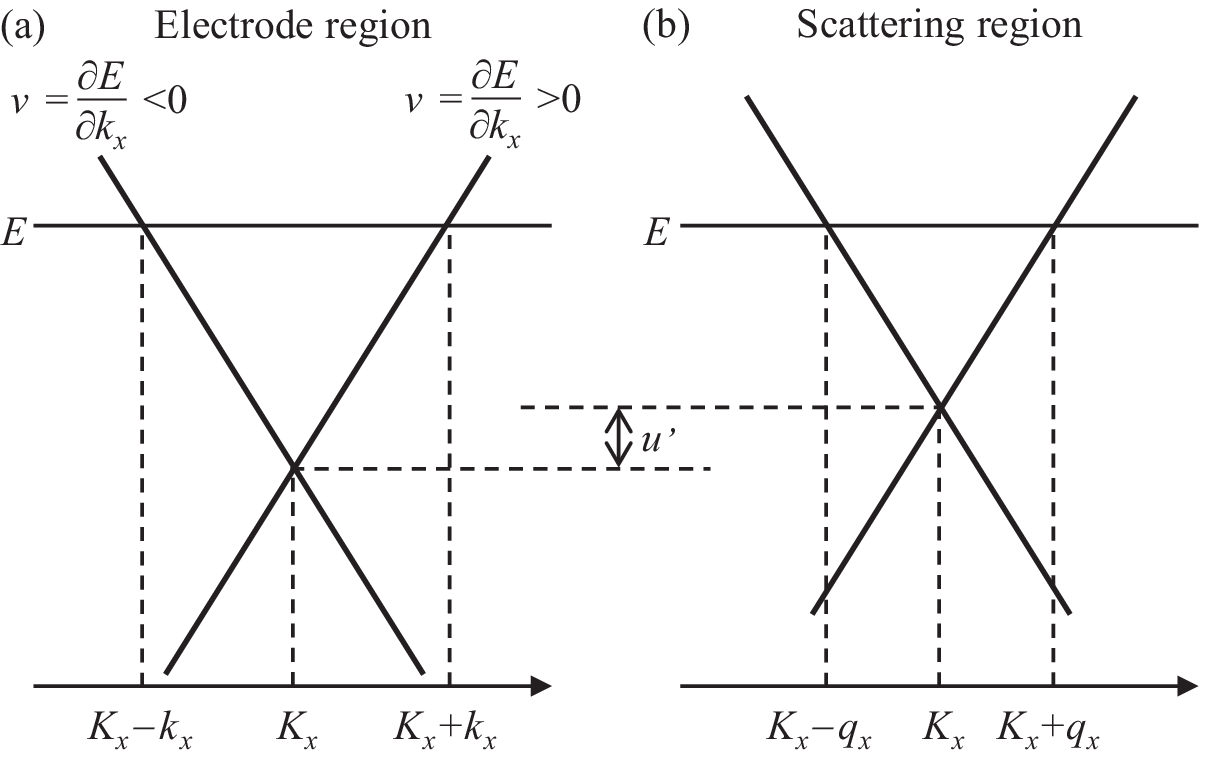}
\caption{Band structures of (a) electrode and (b) scattering regions near $K$ point. The case that $E > 0$ and $u^\prime > 0$ with $u^\prime$ being the potential of the scattering region is shown.}
\label{fig:A3}
\end{center}
\end{figure}

Next, the computational system is divided into three regions using rectangular cells as shown in Fig.~\ref{fig:A2}: left electrode region ($u = 0$), scattering region [$u = u^\prime(\ne 0$)], and right electrode region ($u = 0$). In valley-dependent transport calculations, we assume that incident electron waves are injected from the left electrode with the energy near the Fermi level. When the graphene is calculated using a rectangular cell, the $\mathbf{K}$ and $\mathbf{K}^\prime$ points lie on the $k_x$ axis ($K_y=0$) and $\mathbf{K}=-\mathbf{K}^\prime$. The components of a two-dimensional spinor,
\begin{eqnarray}
\Psi(\mathbf{r}_j)=
\left(
\begin{array}{c}
\psi_A(\mathbf{r}_j) \\
\psi_B(\mathbf{r}_j)
\end{array}
\right), \nonumber
\end{eqnarray}
which corresponds to the amplitude of the wave function in each of two sites that build up the BHC, are introduced. Here, $\mathbf{r}_j$ is the position of the $j$-th atom. By using $\phi_{\mp 1}^{K^\prime}=\phi_{\pm 1}^K(=\phi_\pm)$ when $K_y=0$, we express the scattering wave function. As shown in Fig.~\ref{fig:A3}(a), the scattering wave function at the boundary between the scattering and left electrode regions is
\begin{equation*}
\Psi_L(\mathbf{r}_j)=\mathrm{e}^{i (\mathbf{K} + \mathbf{k}) \cdot \mathbf{r}_j} \phi_+ + r_+ \mathrm{e}^{- i (\mathbf{K} + \mathbf{k}) \cdot \mathbf{r}_j} \phi_+ + r_- \mathrm{e}^{i (\mathbf{K} - \mathbf{k}) \cdot \mathbf{r}_j} \phi_-.
\end{equation*}
In the scattering region, the wave vector is changed to $\mathbf{q}$ from $\mathbf{k}$ because $u \ne 0$, as shown in Fig.~\ref{fig:A3}(b). The scattering wave function in the scattering region is written as
\begin{equation*}
\Psi_C(\mathbf{r}_j)=c_+^+ \mathrm{e}^{i (\mathbf{K} + \mathbf{q}) \cdot \mathbf{r}_j} \phi_+ + c_-^+ \mathrm{e}^{i (\mathbf{K} - \mathbf{q}) \cdot \mathbf{r}_j} \phi_- + c_+^- \mathrm{e}^{- i (\mathbf{K} + \mathbf{q}) \cdot \mathbf{r}_j} \phi_+ + c_-^- \mathrm{e}^{- i (\mathbf{K} - \mathbf{q}) \cdot \mathbf{r}_j} \phi_-.
\end{equation*}
The scattering wave function at the boundary between the scattering and right electrode regions is is written as
\begin{equation*}
\Psi_R(\mathbf{r}_j)=t_+ \mathrm{e}^{i (\mathbf{K} + \mathbf{k}) \cdot \mathbf{r}_j} \phi_+ + t_- \mathrm{e}^{- i (\mathbf{K} - \mathbf{k}) \cdot \mathbf{r}_j} \phi_-.
\end{equation*}
The matching conditions $\Psi_L=\Psi_C$ and $\Psi_L^\prime=\Psi_C^\prime$ at the boundary between the left electrode and scattering regions yield
\begin{eqnarray*}
\mathrm{e}^{i (\mathbf{K} + \mathbf{k}) \cdot \mathbf{r}_j} \phi_+ + r_+ \mathrm{e}^{- i (\mathbf{K} + \mathbf{k}) \cdot \mathbf{r}_j} \phi_+ + r_- \mathrm{e}^{i (\mathbf{K} - \mathbf{k}) \cdot \mathbf{r}_j} \phi_- \\
= c_+^+ \mathrm{e}^{i (\mathbf{K} + \mathbf{q}) \cdot \mathbf{r}_j} \phi_+ + c_-^+ \mathrm{e}^{i (\mathbf{K} - \mathbf{q}) \cdot \mathbf{r}_j} \phi_- + c_+^- \mathrm{e}^{- i (\mathbf{K} + \mathbf{q}) \cdot \mathbf{r}_j} \phi_+ + c_-^- \mathrm{e}^{- i (\mathbf{K} - \mathbf{q}) \cdot \mathbf{r}_j} \phi_-
\end{eqnarray*}
\begin{eqnarray*}
i (K_x+k_x) \mathrm{e}^{i (\mathbf{K} + \mathbf{k}) \cdot \mathbf{r}_j} \phi_+ - i (K_x+k_x) r_+ \mathrm{e}^{- i (\mathbf{K} + \mathbf{k}) \cdot \mathbf{r}_j} \phi_+ + i (K_x-k_x) r_- \mathrm{e}^{i (\mathbf{K} - \mathbf{k}) \cdot \mathbf{r}_j} \phi_-, \\
= i (K_x+q_x) c_+^+ \mathrm{e}^{i (\mathbf{K} + \mathbf{q}) \cdot \mathbf{r}_j} \phi_+ + i (K_x-q_x) c_-^+ \mathrm{e}^{i (\mathbf{K} - \mathbf{q}) \cdot \mathbf{r}_j} \phi_- \\
- i (K_x+q_x) c_+^- \mathrm{e}^{- i (\mathbf{K} + \mathbf{q}) \cdot \mathbf{r}_j} \phi_+ - i (K_x-q_x) c_-^- \mathrm{e}^{- i (\mathbf{K} - \mathbf{q}) \cdot \mathbf{r}_j} \phi_-.
\end{eqnarray*}
By multiplying $\phi_+$ and $\phi_-$ on both sides, we have
\begin{eqnarray}
\label{eqn:match0}
1 + r_+ \mathrm{e}^{- 2i (\mathbf{K}+\mathbf{k}) \cdot \mathbf{r}_j} &=& c_+^+ \mathrm{e}^{- i (\mathbf{k}-\mathbf{q}) \cdot \mathbf{r}_j} + c_+^- \mathrm{e}^{- i (2 \mathbf{K}+\mathbf{k}+\mathbf{q}) \cdot \mathbf{r}_j}, \\
\label{eqn:match1}
1 - r_+ \mathrm{e}^{- 2i (\mathbf{K}+\mathbf{k}) \cdot \mathbf{r}_j} &=& \alpha c_+^+ \mathrm{e}^{- i (\mathbf{k}-\mathbf{q}) \cdot \mathbf{r}_j} - \alpha c_+^- \mathrm{e}^{- i (2 \mathbf{K}+\mathbf{k}+\mathbf{q}) \cdot \mathbf{r}_j}, 
\label{eqn:match2}
\end{eqnarray}
\begin{eqnarray}
r_- &=& c_-^+ \mathrm{e}^{i (\mathbf{k} - \mathbf{q}) \cdot \mathbf{r}_j} + c_-^- \mathrm{e}^{- i (2 \mathbf{K}-\mathbf{k}-\mathbf{q}) \cdot \mathbf{r}_j}, \\
\label{eqn:match3}
r_- &=& \beta c_-^+ \mathrm{e}^{i (\mathbf{k} - \mathbf{q}) \cdot \mathbf{r}_j} - \beta c_-^- \mathrm{e}^{- i (2 \mathbf{K}-\mathbf{k}-\mathbf{q}) \cdot \mathbf{r}_j},
\end{eqnarray}
because $\langle \phi_\pm|\phi_\pm \rangle=1$ and $\langle \phi_\pm|\phi_\mp \rangle=0$. Here,
\begin{eqnarray}
\alpha= \frac{K_x + q_x}{K_x + k_x}, \nonumber \\
\beta = \frac{K_x - q_x}{K_x - k_x} . \nonumber
\end{eqnarray}
The matching conditions for the boundary between the scattering regions and the right electrode regions are obtained in a similar manner.
\begin{eqnarray}
\label{eqn:match4}
c_+^+ \mathrm{e}^{-i (\mathbf{k} - \mathbf{q}) \cdot \mathbf{r}_j} + c_+^- \mathrm{e}^{- i (2 \mathbf{K}+\mathbf{k}+\mathbf{q}) \cdot \mathbf{r}_j} &=& t_+, \\
\label{eqn:match5}
\alpha c_+^+ \mathrm{e}^{-i (\mathbf{k} - \mathbf{q}) \cdot \mathbf{r}_j} - \alpha c_+^- \mathrm{e}^{- i (2 \mathbf{K}+\mathbf{k}+\mathbf{q}) \cdot \mathbf{r}_j} &=& t_+, \\
\label{eqn:match6}
c_-^+ \mathrm{e}^{- i (\mathbf{k} + \mathbf{q}) \cdot \mathbf{r}_j} + c_-^- \mathrm{e}^{- i (2 \mathbf{K}+\mathbf{k} - \mathbf{q}) \cdot \mathbf{r}_j} &=& t_- \mathrm{e}^{- 2 i \mathbf{K} \cdot \mathbf{r}_j}, \\
\label{eqn:match7}
- \beta c_-^+ \mathrm{e}^{- i (\mathbf{k} + \mathbf{q}) \cdot \mathbf{r}_j} + \beta c_-^- \mathrm{e}^{- i (2 \mathbf{K}+\mathbf{k} - \mathbf{q}) \cdot \mathbf{r}_j} &=& t_- \mathrm{e}^{- 2 i \mathbf{K} \cdot \mathbf{r}_j}.
\end{eqnarray}
By solving simultaneous Eqs.~(\ref{eqn:match0}), (\ref{eqn:match1}), (\ref{eqn:match4}), and (\ref{eqn:match5}), we obtain the transmission coefficient $t_+$ for the waves with the same wave vector with the incident wave. On the other hand, the coefficient $t_-$ for the waves with the different wave vector to the incident wave is given by Eqs.~(\ref{eqn:match2}), (\ref{eqn:match3}), (\ref{eqn:match6}), and (\ref{eqn:match7}).
\begin{eqnarray}
\left(
\begin{array}{cccc}
 1 & -\mathrm{e}^{i (\mathbf{k} - \mathbf{q}) \cdot \mathbf{r}_j} & -\mathrm{e}^{i (\mathbf{k} + \mathbf{q}) \cdot \mathbf{r}_j} &  0 \\
 1 & -\beta \mathrm{e}^{i (\mathbf{k} - \mathbf{q}) \cdot \mathbf{r}_j} & \beta \mathrm{e}^{i (\mathbf{k} + \mathbf{q}) \cdot \mathbf{r}_j} &  0 \\
 0 & -\mathrm{e}^{- i (\mathbf{k} + \mathbf{q}) \cdot \mathbf{r}_j} & -\mathrm{e}^{- i (\mathbf{k} - \mathbf{q}) \cdot \mathbf{r}_j} &  1 \\
 0 & \beta \mathrm{e}^{- i (\mathbf{k} + \mathbf{q}) \cdot \mathbf{r}_j} & -\beta \mathrm{e}^{- i (\mathbf{k} - \mathbf{q}) \cdot \mathbf{r}_j} & 1
\end{array}
\right)
\left(
\begin{array}{c}
 r_-  \\
 c_-^+  \\
 c_-^- \mathrm{e}^{- 2i \mathbf{K} \cdot \mathbf{r}_j} \\
 t_- \mathrm{e}^{- 2i \mathbf{K} \cdot \mathbf{r}_j} 
\end{array}
\right)=0 \nonumber
\end{eqnarray}
Since the determinant of the coefficient matrix does not vanish, $t_-=0$, indicating that the intervalley transition does not occur. This result indicates that when the potentials at the A and B sites of the BHC are equal, the scattering problems for the $\mathbf{K}$ and $\mathbf{K}^\prime$ channels can be solved separately. When graphene with the SW defects is divided into the LBHC including several atoms in one cell as shown by blue dashed lines in Fig.~\ref{fig:1}, we can set up the tight-binding Hamiltonian using molecular orbitals instead of atomic orbitals using the LBHC. The transmission probabilities for the $\mathbf{K}^\prime$ ($\mathbf{K}$) channel vanish when electrons are injected into the $\mathbf{K}$ ($\mathbf{K}^\prime$) channel because the determinant of the coefficient matrix for the scattering problem is not zero. 

In the case that $u_A \ne u_B$, the coefficients are obtained by the simultaneous Eqs.~(\ref{eqn:match0})--(\ref{eqn:match7}) because $\langle \phi_\pm|\phi_\mp \rangle \ne 0$ in the scattering region. Therefore, there are no guarantees that $t_-=0$. If the impurities or defects break the symmetry of the BHC, the scattering problems for the $\mathbf{K}$ and $\mathbf{K}^\prime$ channels are not independent. Since the symmetry of the LBHC is destroyed in the S$_2$ structure, the intervalley transition occurs in the S$_2$ structure.

\subsection{Tight-binding calculation}
\begin{figure}[htb]
    \begin{center}
    \includegraphics[width=1.0\textwidth]{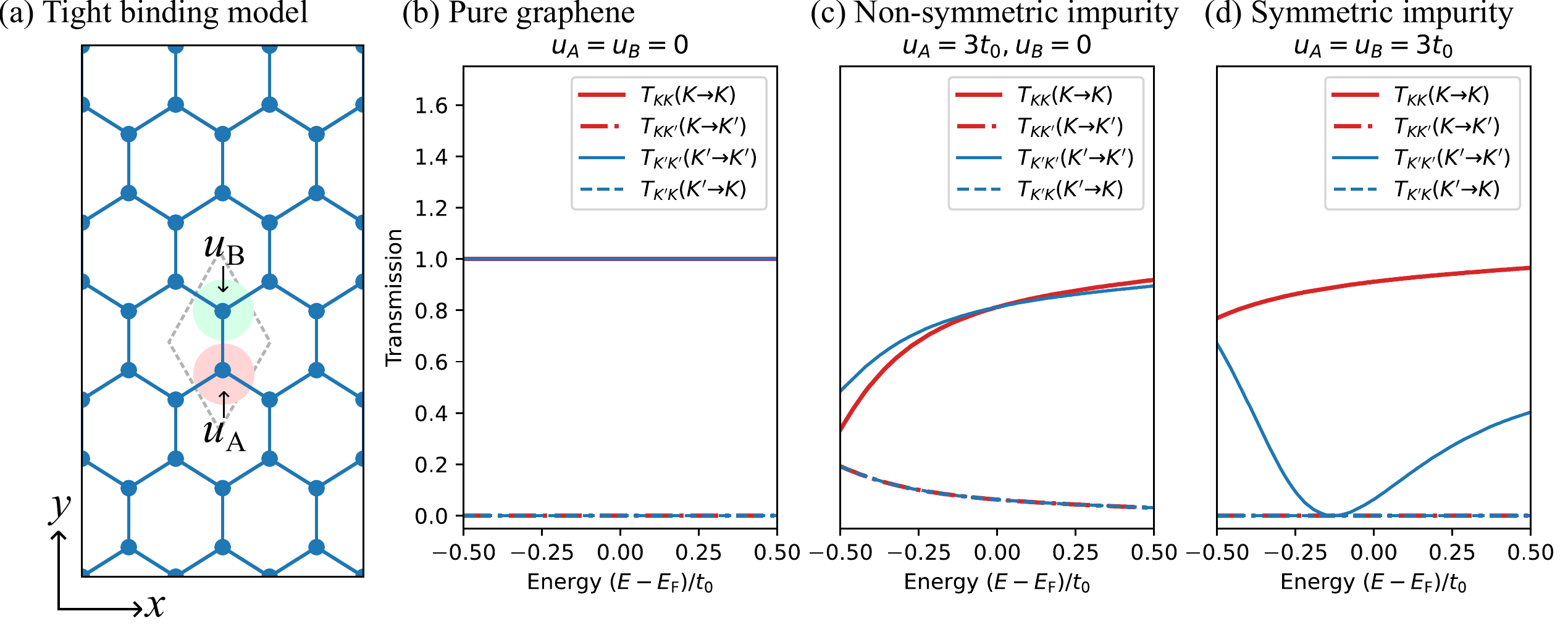}
    \caption{
    (a) Schematic illustration of the tight-binding model: blue circles represent atomic sites, and blue lines indicate nearest-neighbor sites. The large red and green circles show appropriately chosen A and B sites, respectively. $u_\mathrm{A}$ and $u_\mathrm{B}$ are on-site potentials added on A and B sites.
    (b) Calculated channel-resolved transmission spectra for pure graphene.
    (c) Spectra with added repulsive potential only at the A site.
    (d) Spectra with potentials added symmetrically at both A and B sites.
    }
    \label{fig:tbscatter}
    \end{center}
\end{figure}
To ensure the explanation in the preceding subsection, we perform numerical calculations using a simplified tight-binding model; as illustrated in Fig.~\ref{fig:tbscatter}(a), we consider a graphene tight-binding model accounting for only nearest-neighbor transitions with a coefficient of $-t_0$.
The detailed formulation of our theoretical model is provided in \ref{sec:tight_binding}.
For the bulk graphene, the channel-resolved transmission spectra are shown in Fig.~\ref{fig:tbscatter}(b); the transmission spectra of the $\mathbf{K}$ and $\mathbf{K}^\prime$ channels ($T_{KK}$ and $T_{K'K'}$) are unity and symmetric.
Next, we consider the incorporation of non-symmetric impurity sites.
We add a potential $u_\mathrm{A}$ at a single A-site [red circle in Fig.~\ref{fig:tbscatter}(a)], which breaks the symmetry of the LBHC.
The transmission spectra are displayed in Fig.\ref{fig:tbscatter}(b). Near the Fermi level ($E \sim E_\mathrm{F} \sim 0$), valley-polarized conductivity emerges, and the intervalley-transition components exhibit finite values.
Subsequently, we introduce an additional potential $u_\mathrm{B}$ [blue circle in Fig.\ref{fig:tbscatter}(a)] to restore the symmetry.
As depicted in Fig.~\ref{fig:tbscatter}(c), the intervalley transition is prohibited in this case, which reproduces the behavior observed in the S$_1$ structure.
This demonstration explains that intervalley transition is depending on the symmetry of the LBHC of defects.
Thus, it is fair to conclude that the intervalley transition is perfectly suppressed as long as the symmetry of the LBHC is preserved.

\subsection{Graphene blister}
\label{subsec:Graphene blister}
\begin{figure}[htb]
\begin{center}
\includegraphics{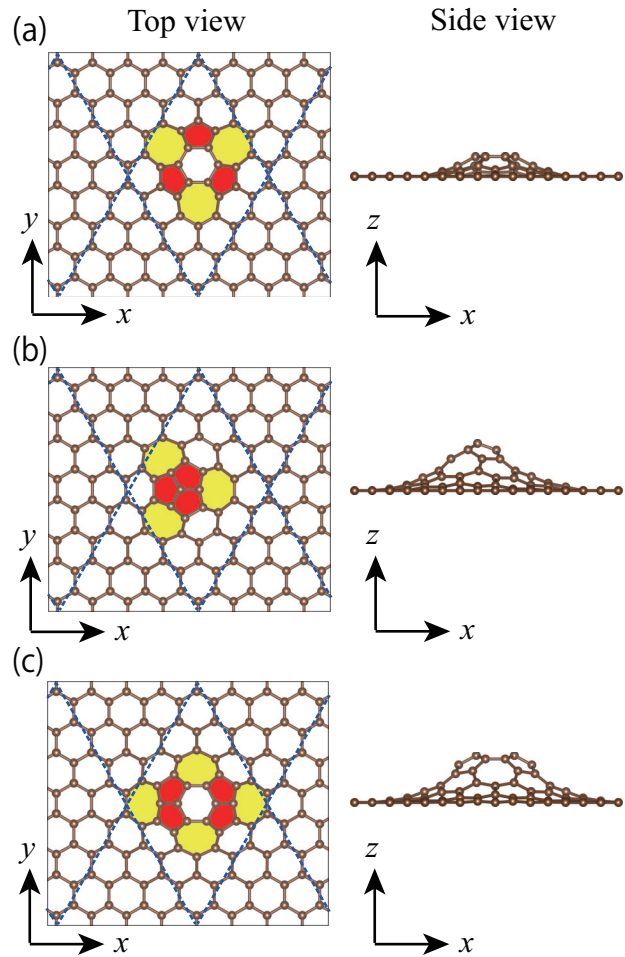}
\caption{Atomic structures of scattering regions of (a) C$_2$, (b) C$_4$, and (c) C$_6$ structures. Blue dotted lines indicate the LBHC containing several carbon atoms.}
\label{fig:6}
\end{center}
\end{figure}

\begin{figure}[htb]
\begin{center}
\includegraphics{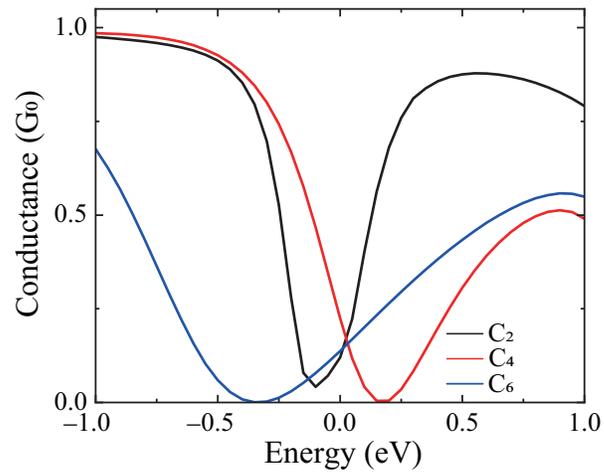}
\caption{Conductance spectra of graphene blisters.}
\label{fig:7}
\end{center}
\end{figure}

\begin{figure}[htb]
\begin{center}
\includegraphics{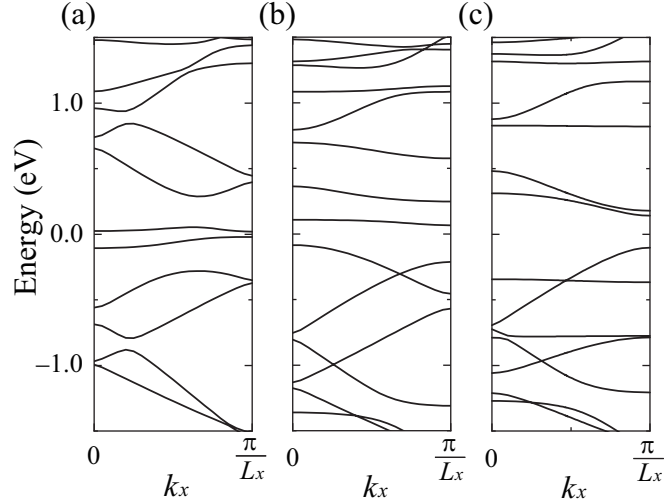}
\caption{Electronic band structures of (a) C$_2$, (b) C$_4$, and (c) C$_6$ structures.}
\label{fig:8}
\end{center}
\end{figure}

\begin{figure*}[htb]
\begin{center}
\includegraphics[width=0.75\textwidth]{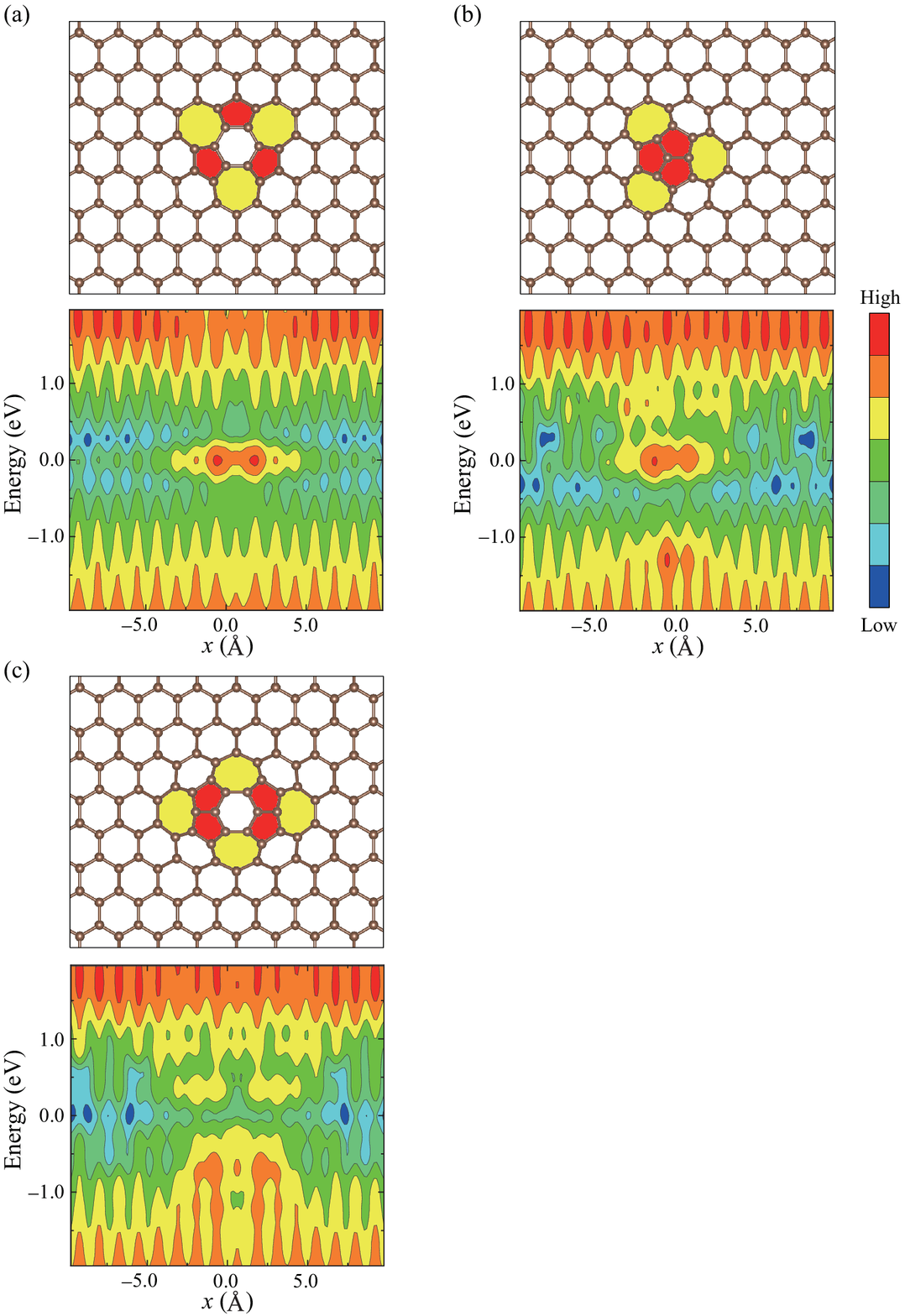}
\caption{LDOSs of (a) C$_2$, (b) C$_4$, and (c) C$_6$ structures. For clarity, structural models are provided on the top of each distribution. Zero energy is chosen as the Fermi level. Each contour represents twice or half the density of the adjacent contours and the lowest contour is 3.78 $\times$ 10$^{-3}$ electron/eV/\AA.}
\label{fig:9}
\end{center}
\end{figure*}

\begin{figure}[htb]
\begin{center}
\includegraphics{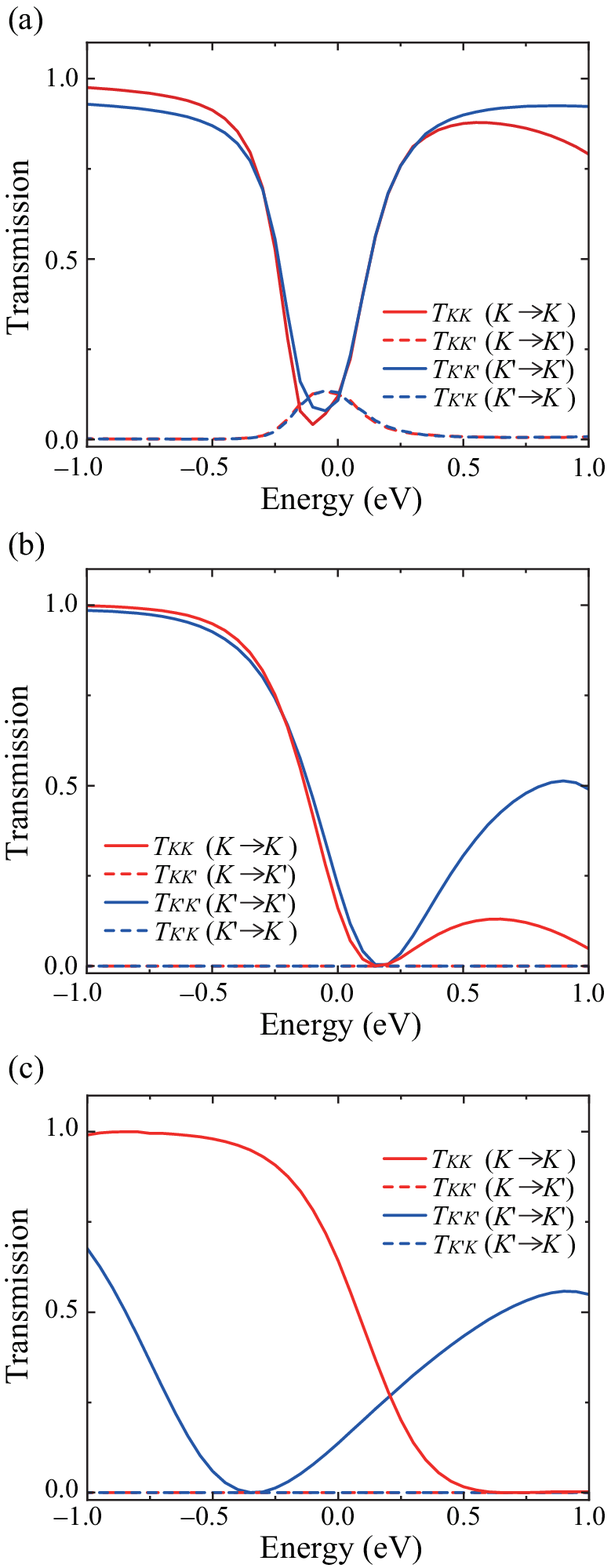}
\caption{Channel-resolved transmission spectra of (a) C$_2$, (b) C$_4$, and (c) C$_6$ structures.}
\label{fig:10}
\end{center}
\end{figure}

Graphene blisters can be generated by adding several carbon atoms to pristine graphene. Electron scattering occurs at the blisters and the symmetry between the A and B sites of the LBHC is destroyed in some cases.  A previous study demonstrated that graphene blisters are created via electron beam irradiation.\cite{NanoLett_14_908} In this subsection, the blisters with up to three additional pairs of carbon atoms are investigated.  Figure~\ref{fig:6} illustrates the atomic structures of scattering regions. The structures with one, two, and three pairs of carbon atoms are named C$_2$, C$_4$, and C$_6$ structures, respectively. The conductance spectra of the graphene blisters are shown in Fig.~\ref{fig:7}. It is found that the electron scattering is larger at the blisters than at the SW defects because the modification of the atomic structure in the blisters is significant. Figures~\ref{fig:8} and ~\ref{fig:9} show the graphene blisters' electronic band structures and LDOSs, respectively, indicating that the states due to the defects lie near the Fermi level and the charge density distributions of the defect state spatially accumulate around the blisters. We plot in Fig.~\ref{fig:10} the channel-resolved transmission spectra. In the C$_2$ structure, since the symmetry between the A and B sites of the LBHC is broken, the intervalley transition occurs but not in the other structures. Furthermore, since the level of electron scattering is high for the blisters, the energy regions where only the electrons in the $\mathbf{K}$ ($\mathbf{K}^\prime$) valley are transmitted while the others are completely reflected appear. These results indicate that the valley-polarized current appears in the blister defects owing to the local application of a gate voltage to the blister region in the same way as in graphene quantum point contact devices \cite{NatPhys.3.172}. 

\section{Conclusion}
\label{sec:conclusion}
We have proposed the application of graphene valley filters using the blister defects by calculating the channel-resolved conductance spectra of various structural defects using DFT techniques. In SW defect structures (S$_1$ and S$_2$), the intervalley transition ($T_{KK'}$) is observed in the S$_2$ structure, whereas no intervalley transition occurs in the S$_1$ structure, although the BHC is distorted in both models. By employing the LBHC containing several carbon atoms and replacing atomic orbitals with molecular orbitals in the tight-binding model, we have proved analytically and numerically that the symmetry breaking of the structural defects induces intervalley transition. 
In addition, we have proposed the universal rule for the blister defects that the symmetry between the A and B sites of the LBHC of the defect plays an important role in the conservation of the valley index. 
It is found that the energies of the states in which resonant scattering occurs on the $\mathbf{K}$ and $\mathbf{K}^\prime$ channel electrons split in the blister defects (C$_2$, C$_4$, and C$_6$) owing to the carbon atoms added. The split of the resonant states will enable us to realize the fully valley-polarized current by the local application of a gate voltage to the blister region.

\ack
This work was partially financially supported by MEXT as part of the “Program for Promoting Researches on the Supercomputer Fugaku” (Quantum-Theory-Based Multiscale Simulations toward the Development of Next-Generation Energy-Saving Semiconductor Devices, JPMXP1020200205) and also supported as part of the JSPS KAKENHI (JP22H05463), JST CREST  (JPMJCR22B4), and JSPS Core-to-Core Program (JPJSCCA20230005). The numerical calculations were carried out using the computer facilities of the Institute for Solid State Physics at The University of Tokyo, the Center for Computational Sciences at University of Tsukuba, and the supercomputer Fugaku provided by the RIKEN Center for Computational Science (Project ID:
hp230175). 
We would like to thank Editage (www.editage.com) for English language editing.

\section*{Data availability statement}
All data that support the findings of this study are included in the article.

\appendix
\section{Tight-binding transport calculation}
\label{sec:tight_binding}
\subsection{Hamiltonian for transport calculation}
\begin{figure}[htb]
    \begin{center}
    \includegraphics[width=0.5\textwidth]{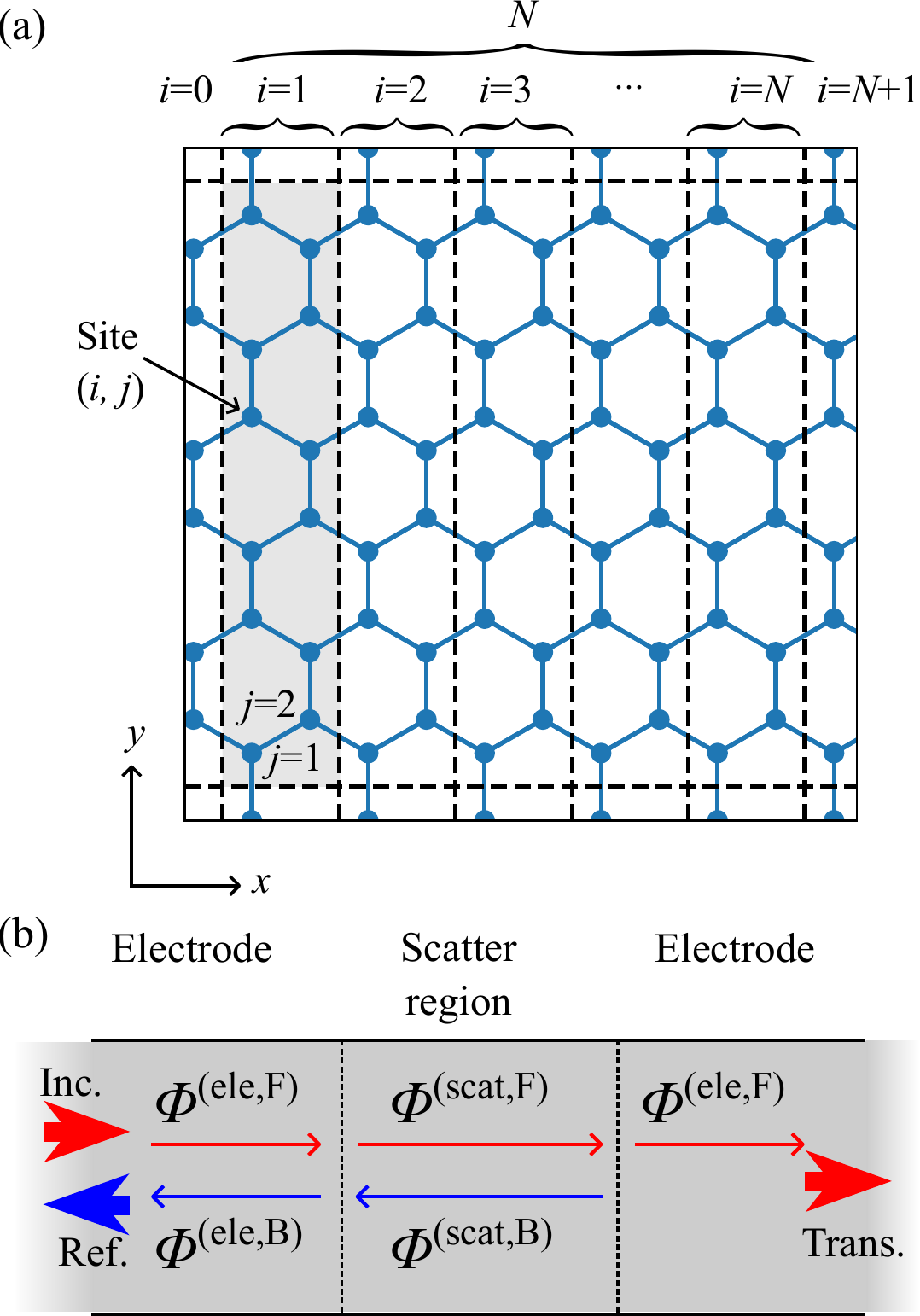}
    \caption{
    Schematic illustration of the computational model: Blue circles represent carbon atom sites, and blue lines indicate the connections between nearest-neighbor sites. Rectangular units (in gray) are arranged along the $x$ axis (zigzag direction) and are denoted by index $i$.
    }
    \label{fig:tight_binding}
    \end{center}
\end{figure}

In this work, the tight-binding model is utilized to model the transport properties of graphene with structural defects.
The schematic illustration is shown in Fig.~\ref{fig:tight_binding}.
The supercell is composed of $N$ rectangular units (the gray area), represented by index $i$; each unit contains $M$ atomic sites, represented by index $j$; every site is denoted by the pair of indices $(i, j)$.
Each unit is arranged along the $x$ axis, corresponding to the zigzag direction.

In the tight-binding model, the wave function $\psi$ is described as below:
\begin{eqnarray}
    \psi_{\mathbf{k}}(\mathbf{r})=
    \sum_{i,j}
    c_{i,j} \chi(\mathbf{r}-\mathbf{r}_{i, j}) e^{i \mathbf{k} \mathbf{r}_{i, j}}
    \;,
\end{eqnarray}
where $\chi$ represents the atomic wave function corresponding to the $\pi$-orbital of the carbon atom, $c_{i,j}$ and $\mathbf{r}_{i,j}$ are the expansion coefficient and position of the site $(i, j)$, respectively, and $\mathbf{k}$ is the Bloch wavevector along the $y$ axis.
The Hamiltonian matrix $\hat{\mathcal{H}}$ has finite values only for the diagonal elements and off-diagonal elements corresponding to the nearest-neighbor (NN) sites:
\begin{eqnarray}
    \int \;
    \chi(\mathbf{r}-\mathbf{r}')
    \; \hat{\mathcal{H}}
    \chi(\mathbf{r}-\mathbf{r}'')
    \; \mathrm{d}\mathbf{r}
    =
    \left\{\begin{array}{cc}
        u(\mathbf{r}') & (\mathbf{r}' = \mathbf{r}'') \\
        -t_0 & ((\mathbf{r}' , \mathbf{r}'')  \in \mathrm{NN}) \\
        0 & \mathrm{otherwise}
    \end{array}\right.
    \;.
\end{eqnarray}
The eigenvalue problem for the Schr\"odinger equation $\hat{\mathcal{H}} \psi = E \psi$ can be expressed as the following matrix form:
\begin{eqnarray}
\label{eqn:schroedinger-b3}
    \left(\begin{array}{cccccc}
        \ddots &
        \vdots &
        \vdots &
        \vdots &
        \vdots &
        \\
        \cdots &
        \hat{{\mathbf{H}}}_{0,0} &
        \hat{{\mathbf{B}}}_{0,1} &
        0 &
        0 &
        \cdots 
        \\
        \cdots &
        \hat{{\mathbf{B}}}_{1,0} &
        \hat{{\mathbf{H}}}_{1,1} &
        \hat{{\mathbf{B}}}_{1,2} &
        0 &
        \cdots
        \\
        \cdots &
        0 &
        \hat{{\mathbf{B}}}_{2,1} &
        \hat{{\mathbf{H}}}_{2,2} &
        \hat{{\mathbf{B}}}_{2,3} &
        \cdots
        \\
        \cdots &
        0 &
        0 &
        \hat{{\mathbf{B}}}_{3,2} &
        \hat{{\mathbf{H}}}_{3,3} &
        \cdots
        \\
         &
        \vdots &
        \vdots &
        \vdots &
        \vdots &
        \ddots
    \end{array}\right)
    \left(\begin{array}{c}
        \vdots \\
        \Psi_{0} \\
        \Psi_{1} \\
        \Psi_{2} \\
        \Psi_{3} \\
        \vdots
    \end{array}\right)
    &=&
    E
    \left(\begin{array}{c}
        \vdots \\
        \Psi_{0} \\
        \Psi_{1} \\
        \Psi_{2} \\
        \Psi_{3} \\
        \vdots
    \end{array}\right)
    \;.
\end{eqnarray}
For convenience, we employ the block form; 
$\Psi_i \equiv (c_{i,1}, c_{i,2}, \cdots, c_{i,j}, \cdots)^\mathrm{T}$ is a block vector consisting of coefficients belonging to the sites in the $i$-th unit.
$\hat{\mathbf{H}}_{i, i}$ and $\hat{\mathbf{B}}_{i,i\pm1}$ represent the $M$-dimensional block Hamiltonian matrices for the same unit and the neighboring units, respectively.
All other matrix elements are zero. The Schr\"odinger equation (\ref{eqn:schroedinger-b3}) is truncated as
\begin{eqnarray*}
    \left(\begin{array}{ccccc}
        E \hat{I} - \hat{{\mathbf{H}}}_{1,1} &
        - \hat{{\mathbf{B}}}_{1,2} &
        0 &
        \cdots &
        0
        \\
        - \hat{{\mathbf{B}}}_{2,1} &
        E \hat{I} - \hat{{\mathbf{H}}}_{2,2} &
        - \hat{{\mathbf{B}}}_{2,3} &
        \cdots &
        0
        \\
        0 &
        - \hat{{\mathbf{B}}}_{3,2} &
        E \hat{I} - \hat{{\mathbf{H}}}_{3,3} &
        \cdots &
        0 
        \\
        &
        &
        &
        \ddots &
        \\
        0 &
        0 &
        0 &
        \cdots &
        E \hat{I} - \hat{\mathbf{H}}_{N,N}
    \end{array}\right)
    \left(\begin{array}{c}
        \Psi_1 \\
        \Psi_2 \\
        \Psi_3 \\
        \vdots \\
        \Psi_{N}
    \end{array}\right)
    &=&
    \left(\begin{array}{c}
        \hat{\mathbf{B}}_{1,0} \Psi_{0}  \\
        0 \\
        0 \\
        \vdots \\
        \hat{\mathbf{B}}_{N,N+1} \Psi_{N+1}
    \end{array}\right)
\end{eqnarray*}
with $\hat{I}$ being an $M$-dimensional identity matrix. Here, we introduce the Green's function $\hat{\mathcal{G}}$ as the inverse matrix of the matrix on the left-hand side.
By solving the above, we obtain the following equations.
\begin{eqnarray}
    \Psi_1 =& {\hat{G}}_{1,1} \hat{\mathbf{B}}_{1,0} \Psi_{0} +  {\hat{G}}_{1,N} \hat{\mathbf{B}}_{N,N+1} \Psi_{N+1}
    \label{eq:green1}
    \;,
    \\
    \Psi_{N} =& {\hat{G}}_{N,1} \hat{\mathbf{B}}_{1,0} \Psi_{0} +  {\hat{G}}_{N,N} \hat{\mathbf{B}}_{N,N+1} \Psi_{N+1}
    \label{eq:green2}
    \;,
\end{eqnarray}
where $\hat{G}_{m,n}$ represents the $M$-dimensional $(m, n)$ block matrix elements of the Green's function $\hat{\mathcal{G}}$.
As shown in Fig.~\ref{fig:tight_binding}, the computational model for transport-property calculation consists of three domains: the scattering region is sandwiched by two electrode regions.
In the electrode regions, we assume the Hamiltonian of pure graphene and the wave functions $\Psi$ in the electrodes without any scatterings are denoted as the generalized Bloch waves $\Phi^\mathrm{(elec)}$. In the scattering region, we introduce an on-site potential to account for the effects of defects in the Hamiltonian and the wave functions $\Psi$ with scatterings are denoted as the scattering wave functions $\Psi^\mathrm{(scat)}$.

\subsection{Generalized Bloch wave function in electrode regions}
For the pure graphene, periodic boundary conditions of length $N=1$ are imposed, and the wave function is required to satisfy generalized Bloch condition: $\Phi^\mathrm{(elec)}_{2} = \lambda \Phi^\mathrm{(elec)}_1$ and $\Phi^\mathrm{(elec)}_{0} = \lambda^{-1} \Phi^\mathrm{(elec)}_1 $, where $\lambda$ is a Bloch phase factor.
By substituting the Bloch condition to Eq.~(\ref{eqn:schroedinger-b3}), we obtain a quadratic eigenvalue problem providing the generalized Bloch wave functions in the electrode regions:
\begin{equation}
\label{eqn:QEP}
\lambda^{-1} \hat{\mathbf{B}}_{1,0} \Phi^\mathrm{(elec)}_1 + \left[ E \hat{I} - \hat{{\mathbf{H}}}_{1,1} \right] \Phi^\mathrm{(elec)}_1 + \lambda \hat{\mathbf{B}}_{1,2} \Phi^\mathrm{(elec)}_1 = 0.
\end{equation}
Equation (\ref{eqn:QEP}) can be solved by replacing to the generalized eigenvalue problem \cite{PhysRevB.98.195422}.

\subsection{Scattering wave function}
The generalized Bloch wave functions in the electrode regions obtained from Eq.~(\ref{eqn:QEP}) can be classified into forward-propagating (F) waves and backward-propagating (B) waves:
$\Phi^{(\mathrm{elec},\mathrm{F}, l)}$, $\Phi^{(\mathrm{elec},\mathrm{B}, l)}$ with channel index $l$.
We determine the scattering wave function in the scattering region $\Psi_i^\mathrm{(scat)}$.
The continuity between the wave function in the electrode regions and that in the scattering region is expressed as 
\begin{eqnarray}
    \Psi^\mathrm{(scat)}_{0} =& \Phi^{(\mathrm{elec},\mathrm{F},{l^\prime})} + \sum_l r_l \Phi^{(\mathrm{elec},\mathrm{B},l)} \;, \\
    \Psi^\mathrm{(scat)}_{1} =&  \lambda_{l^\prime} \Phi^{(\mathrm{elec},\mathrm{F},{l^\prime})} + \sum_l r_l \lambda_l \Phi^{(\mathrm{elec},\mathrm{B},l)} \;, \\
    \Psi^\mathrm{(scat)}_{N} =& \sum_l t_l \lambda_l^{-1} \Phi^{(\mathrm{elec},\mathrm{F},l)} \;, \\
    \Psi^\mathrm{(scat)}_{N+1} =& \sum_l t_l \Phi^{(\mathrm{elec},\mathrm{F},l)} 
    \;,
\end{eqnarray}
where $r_l$ and $t_l$ are the coefficients of reflection and transmitted waves, respectively.
By combining the above condition with Eqs.~(\ref{eq:green1}) and (\ref{eq:green2}) for $\Psi^\mathrm{(scat)}$, we obtain the coefficients $r_l$, and $t_l$.
The conductance can be obtained from the following relation.
\begin{eqnarray}
    G = G_0 \sum_l |t_l|^2 \frac{v_{\mathrm{g},l}}{v_{\mathrm{g}, {l^\prime}}}
    \;,
\end{eqnarray}
where $v_{\mathrm{g},l}$ is the group velocity for channel $l$ and $G_0$ is the conductance quantum unit: 
$G_0 = {2e^2}/{h} \approx 1 / (12.9 \: \mathrm{k}\Omega)$. 
In addition, following a similar procedure to Eq.~(\ref{eq:separate}), it becomes possible to calculate the valley-component of conductivity:
$T_{KK}$, $T_{K'K}$, $T_{KK'}$ and $T_{K'K'}$.

\section*{References}
\bibliographystyle{iopart-num}
\bibliography{ref}

\end{document}